Collapse of the Fe-vacancy order and successive phase transitions in superconducting $K_xFe_{2-y}Se_2$ ($0.7 \leq x \leq 0.8$, $0.2 \leq y \leq 0.3$)

J. Q. Li[1]*, Y. J. Song[1], H. X. Yang[1,] Z. Wang[1], H. L. Shi[1], G. F. Chen[2], Z.W. Wang[1], Z.Chen[1] and H. F. Tian[1].

.

[1] Beijing National Laboratory for Condensed Matter Physics, Institute of Physics, Chinese Academy of Sciences, Beijing 100190, P. R. China

[2] Department of Physics, Renmin University of China, Beijing 100872, P. R. China

The Fe-based superconducting $K_xFe_{2-y}Se_2$ ($0.6 \leq x \leq 1$, $0.2 \leq y \leq 0.4$) compounds, unlike the well-known $RFe_2As_2$ (R = Ba，Sr，Ca) superconductors, contain complex structural features and notable physical properties, such as the Fe-vacancy ordering, multi-superconducting transitions and the phase separation[1]. Recent experimental studies also suggested the presence of remarkable interplay among the Fe-vacancy order and the antiferromagnetic structure[2]. Here we demonstrate that the well-characterized superconducting $K_xFe_{2-y}Se_2$ ($0.6 \leq x \leq 0.8$, $0.2 \leq x \leq 0.3$) samples contain complex microstructure features and undergo successive phase transitions at low temperatures. In-situ TEM observations on a number of the $K_xFe_{2-y}Se_2$ superconductors demonstrated the presence of a remarkable collapse of the Fe-vacancy order above the **superconducting transition, as a result, the superconducting phase actually adopts a tetragonal structure without the Fe-vacancy ordering. Moreover, our analysis at the low temperatures suggests that the superconductors likely adopt a Fe-deficient structure with composition of $K_{0.75}Fe_{2-y}Se_2$. These results are important not only for the further optimization of superconducting phase in present system but also for understanding the mechanism of the Fe-based superconductivity.**

*Corresponding author. E-mail: LJQ@aphy.iphy.ac.cn

Following the discovery of superconductivity up to 26 K in LaFeAsO$_{0.89}$F$_{0.11}$ [3], a number of Fe-based superconductors with different structural features have been obtained, such as the '111' and the '122' type structural phases[4-6]. In particular, the SmFeAsO$_{0.85}$ compound shows the superconducting transition at the critical temperature as high as 55 K [7]. The theoretical and experimental investigations on the Fe-based systems demonstrate that these superconductors, as observed in the high-Tc cuprates [8], exhibit remarkable interplay between magnetism and superconductivity[9]. In addition to the FeAs-layer systems, superconductivity has also been studied in the related compound of FeSe$_{1-x}$ (x~0.2) [10], which could exhibit a visible increase of $T_c$ up to 37 K under a pressure of 6 GPa [11]. Very recently, superconductivity at about 30K was reported in the FeSe-layer compound KFe$_2$Se$_2$ [12], then the superconductivity have been also observed in (Tl,K)Fe$_{2-y}$Se$_2$ and Rb$_x$Fe$_{2-y}$Se$_2$ with Tc above 30 K[13-15]. These materials in general show up complex microstructure and magnetic properties, moreover, the coexistence of the superconductivity and anti-ferromagnetism, as a main focus in the recent debated issues, has been extensively discussed in both experimental and theoretical studies. [2,16-18].

Both single crystal and polycrystalline samples were used in present study. The single crystals were grown by the Bridgeman method as reported previously in ref. [19]. Polycrystalline samples were prepared by a two-step method [20]. Resistivity below 400 K was measured using the Quantum Design PPMS-9 by a standard four-point probe method. Powder X-ray diffraction is performed on a Bruker AXS D8 Advance diffractometer equipped with TTX 450 holder working in the temperature rang of 80 K ~ 700 K. Specimens for TEM observations were prepared by peeling off a very thin sheet of a thickness around several tens microns from the single crystal and milling by Ar ion under low temperature. Microstructure and phase transition investigations were performed on a FEI Tecnai-F20 transmission electron microscope (TEM) equipped with a double-tilt cooling holder down to 17K.

The Fe-vacancy ordering as an important phenomenon has been extensively investigated in $K_xFe_{2-y}Se_2$ compounds due to its specific significance for understanding of the antiferromagnetic behavior and superconducting mechanism [2, 22]. In order to perform a systematic study of the microstructure and structural phase transitions in this layered system, we selected a few well-characterized $K_xFe_{2-y}Se_2$ ($0.6 \leq x \leq 0.8$, $0.2 \leq y \leq 0.3$) samples. In Fig.1, we show the temperature-dependences of resistivity for two typical superconducting samples (one polycrystalline and the other single crystal sample) which were used mostly in our experimental studies, they both have sharp superconducting transitions. Inset shows the magnetic susceptibility for the single crystal, and the superconducting volume fraction is the estimated to be larger than 90 %. On the other hand, a semiconducting $K_{0.8}Fe_{1.6}Se_2$ (so called the $K_2Fe_4Se_5$ phase) polycrystalline sample [20] containing the well-defined Fe-vacancy order was also used in our study for comparison. This sample has an antiferromagnetic structure in association with the Fe-vacancy order along [130] direction as reported in previous literature [1]. The inserted image shows the superstructure model containing the Fe-vacancy order in $K_2Fe_4Se_5$.

In Fig. 2, we show the [001] zone-axis electron diffraction patterns as observed by in-situ cooling TEM observations on a well-characterized single crystal of $K_{0.7}Fe_{1.7}Se_2$ with superconducting transition at $T_c \sim 31K$, illustrating modification of the superstructure in the superconducting samples at low temperatures. The main diffraction spots in the diffraction pattern at 300 K (Fig. 2 (a)) can be well indexed with the aforementioned tetragonal unit cell with lattice parameters of a= 3.913 Å and c=14.10 Å and a space group I4/mmm. The most pronounced structural feature demonstrated in present in-situ TEM observation is the disappearance of the superstructure reflections association with the Fe-vacancy order above the superconducting critical transition ($T_c \sim 31K$). As discussed in previous literatures[2, 20], the superstructure spots become visible associated with the Fe-vacancy disorder-to-order transformation at around 560 K following by an antiferromagnetic transition. This Fe-vacancy ordering results in a clear structural modulation going along the [130] or [3-10] directions. The wave vector of this modulation can be written as $q_1=1/5(a* + 3b*)$, and insert of Fig. 1(b) schematically displays a structural model for the Fe-vacancy order which is consistent with the ordering wave vector $q_1$. According to in-situ TEM experimental data obtained from a number of well characterized superconducting samples with compositions of $K_xFe_{2-y}Se_2$ ($0.6 \leq x \leq 0.8$, $0.2 \leq y \leq 0.3$), all superstructure spots become steadily weaker below 200K and then turn out to be invisible below 70K. Hence, we can conclude that the superconducting transition in present system occurs in a tetragonal basic structure instead of the superstructure phase as discussed in previous literatures[2, 18]. Moreover, our observations on the alternation of superstructure also suggest that this structural transition is almost reversible, and no clear hysteretic nature is observed as measured by means of electron and X-ray diffractions. On the other hand, it is also noted in low-temperature diffraction pattern that certain diffuse spots at the systematic (1/2 1/2 0) position become clearly visible at 20K, as shown in Fig. 2(d). These superlattice reflections are likely in correlation with a K-ion order, as discussed in the following context.

A clear view for the temperature dependence of the superstructure in correlation with the Fe-vacancy ordering has been also obtained by the powder X-ray diffraction measurements in the large temperature range from 80K up to 580K. Fig. 3(a) shows a series experimental data taken from a polycrystalline sample with $T_c \sim 27K$. The superstructure reflections as indicated by an arrow are clear visible at $2\theta=14.6^o$ in the diffraction pattern, it can be well-indexed to the Fe-vacancy order along [130] direction with a periodicity of L = 0.62nm. It is remarkable that this superstructure become gradually weaker with lowering temperature from 200K down to 80K, this results are fundamentally in consistence with the in-situ TEM observations as mentioned in above context. On the other hand, the X-ray diffraction patterns obtained at high-temperature range show clearly changes above 450K, and the superstructure peaks become invisible above 550K where the Fe-vacancy order-to-disorder phase transition occurs as reported in ref. [20]. Otherwise, we can also see a few additional peaks from $FeSe_{1-x}$ appearing in high-temperature diffraction patterns as typically indicated by an asterisk, this fact suggests a partial decomposition of the superconducting $K_xFe_{2-y}Se_2$ samples occurs in the Ar environment as used in our X-ray diffraction experiments. Fig. 3(b) shows the normalized intensity of the superstructure peak at 14.6º as a function of the temperature obtained in this superconducting sample. It is remarkable that the intensity of the superstructure shows visible decrease blow 200K as also illustrated in inserted image in Fig. 3(a). In contrast, in-situ cooling measurements by electron and X-ray diffractions on the $K_{0.8}Fe_{1.6}Se_2$ (i.e. the $K_2Fe_4Se_5$ phase) semiconductor demonstrate a rather stable Fe-vacancy order at low temperatures, and no visible changes on either the intensity or position of superlattice reflections were detected at low temperatures. In Fig.3 (b), we also show the experimental data obtained from a $K_2Fe_4Se_5$ sample for comparison, this sample contains remarkable Fe-vacancy order and has a large resistivity as shown in Fig. 1(b).

Let us have a brief discussion on the correlated features between the transport properties and structural changes in the superconducting $K_xFe_{2-y}Se_2$ materials. In Fig. 3(d), we display the temperature dependence of the resistivity for the superconducting sample used in our X-ray measurements. This resistivity profile shows a broad hump as commonly observed in the $AFe_{2-y}Se_2$ ( A=K,Tl, Rb) superconductors [12-15,19]. It is recognizable that the superstructure intensity also shows a hump-like feature in the same temperature range. It also noted that a recent NMR measurement on $Tl_{0.47}Rb_{0.34}Fe_{1.63}Se_2$ superconductor suggest the pseudogap opening at 400 K [21] where the hump-like anomaly occurs in resistivity. These facts suggest visible effects of the Fe-vacancy ordering on physical properties in this superconducting system. Actually, the origin, as well as the driving force trigging collapse of the Fe-vacancy order, could be an essential issue for understanding the structural transitions, the magnetic structure and complex transport properties in this layered system.

The low-temperature crystal structure, in which the superconducting transition occurs, is another critical issue concerned in our investigation. Though the average structure of this low-temperature phase can be assigned to a tetragonal structure as illustrated in Fig. 2(c) and (d) and confirmed by X-ray diffractions, our careful analysis on structural data obtained at 20K also reveal a series of superstructure reflections at systematic (1/2 1/2 0) position. In order to determine the low-temperature crystal structure, we have also performed our TEM observations along the relevant directions. Fig. 4(a) and (b) show the diffraction patterns taken respectively along the [001] and [110] zone axis directions, they both exhibit the additional reflections of the (1/2 1/2 0) superstructure. This low-temperature (1/2 1/2 0) order, in sharp contrast with the Fe -vacancy order, often becomes evidently visible below 50K. In certain crystals, it has a short correlation length along the c-axis as shown in Fig.4 (b). We herein interpret these structural behaviors in terms of the K-ion order which turns to stable at low temperatures. Fig. 4(c) shows a structural model for the superconducting phase with a (1/2 1/2 0) potassium-order within the a-b plane. Theoretical simulations based on this structural model could give rise to clear superstructure spots in good agreement with the experimental ones as illustrated in Fig.4 (d) and (e).

It is believed that the superconducting phase in K-Fe-Se system should have certain deviations from the stoichiometric $KFe_2Se_2$ and even from the nominal compositions of the superconducting samples used in recent studies. According to the low-temperature structural data, the superconducting materials can be assigned into a Fe-deficient structure with chemical composition of $K_{0.75}Fe_{2-y}Se_2$. Recently, we have prepared a series of compounds of with composition of $K_{0.75}Fe_{2-y}Se_2$ ($0 \leq y \leq 0.4$). Structural analysis suggests that the single-phase materials can be obtained for $0.2 \leq y \leq 0.4$. Measurements of transport and magnetic properties reveal that samples with $0.2 \leq y \leq 0.35$ often show sharp superconducting transitions between 25 and 33K. On the other hand, it is known that the Cu-based high-$T_c$ superconductors often show the optimal superconductivity for the density of charge carrier at $n_c$=0.15 [23], our analysis on the $K_{0.75}Fe_{2-y}Se_2$ system suggest that $K_{0.75}Fe_{2-y}Se_2$ (y=0.25) compound have a nominal charge density $n_c = 0.15$, indeed, it shows a sharp superconductivity above 30K. This fact likely suggests that the Fe-based superconductor could have certain essential features in common with the high-Tc cuprates. A further study on the superconductivity, crystal structure and electronic phase diagram of the $K_{0.75}Fe_{2-y}Se_2$ system is in progress.

In summary, $K_xFe_{2-y}Se_2$ ($0.7 \leq x \leq 0.8$, $0.2 \leq x \leq 0.3$) materials often show complex microstructure features and undergo successive phase transitions with lowering temperature. The most important phase transition resulting from collapse of the Fe-vacancy order occurs commonly in superconducting samples at low-temperatures. In-situ cooling measurements on the superstructure suggest that the hump-like anomaly in resistivity is probably in correlation with the defused feature of the Fe-vacancy ordering collapse. Low-temperature structural study demonstrates that superconducting phase has a tetragonal average structure in which the K ions possibly adopts an (1/2 1/2 0) order in the a-b plane but with a short coherence length along c-axis direction. These results are important for the understanding of the rich microstructure phenomena and for further optimization of superconducting properties in present system.

**Acknowledgments**

This work is supported by the National Science Foundation of China, the Knowledge Innovation Project of the Chinese Academy of Sciences, and the 973 projects of the Ministry of Science and Technology of China.

Figure captions

Fig.1. (a) Temperature dependences of electric resistivity for a $K_{0.8}Fe_{1.8}Se_2$ polycrystal and a $K_{0.7}Fe_{1.7}Se_2$ single-crystal superconductors. Inset shows the magnetization measured with $H_{//ab}$= 50Oe for $K_{0.7}Fe0_{1.7}Se_2$. (b) Temperature dependence of resistivity for a $K_{0.8}Fe_{1.6}Se_2$ (the 245-phase) semiconductor, inserted image shows a structural model for the Fe-vacancy order with a modulation of $q_1=a*/5+3b*/5$.

Fig. 2. The [001] zone-axis electron diffraction patterns taken from a superconducting $K_{0.7}Fe0_{1.7}Se_2$ single crystal at low temperatures, the superlattice reflections at systematic (1/5 3/ 5 0) position become invisible below 70K.

Fig.3. (a) X-ray diffraction patterns showing the temperature dependence of superstructure in a superconducting $K_{0.8}Fe_{1.8}Se_2$ sample. Inset shows the intensity alteration of a superstructure peak at $2\theta=14.6^o$ with lowering temperature. (b) Relative intensity of the superstructure in the $K_{0.8}Fe_{1.8}Se_2$ superconductor. The data obtained from the $K_{0.8}Fe_{1.6}Se_2$ semiconductor is also shown for comparison. (c) Temperature dependence of electric resistivity of $K_{0.8}Fe_{1.8}Se_2$ superconductor exhibits a hump-like anomaly in intermediate temperature region comparable with what observed for the superstructure as shown in Fig. 3(b).

Fig.4. Low-temperature structural features at 20K for a $K_{0.75}Fe0_{1.7}Se_2$ superconductor with Tc=31K. Electron diffraction patterns taken along (a) the [001] and (b) the [110] zone-axis direction, illustrating the tetragonal basic structure modulated by a (1/2 1/2 0) order. Theoretical simulation data based on a K-ion ordered state as shown in Fig. 4(c) are presented in Fig. 4(d) and (e).

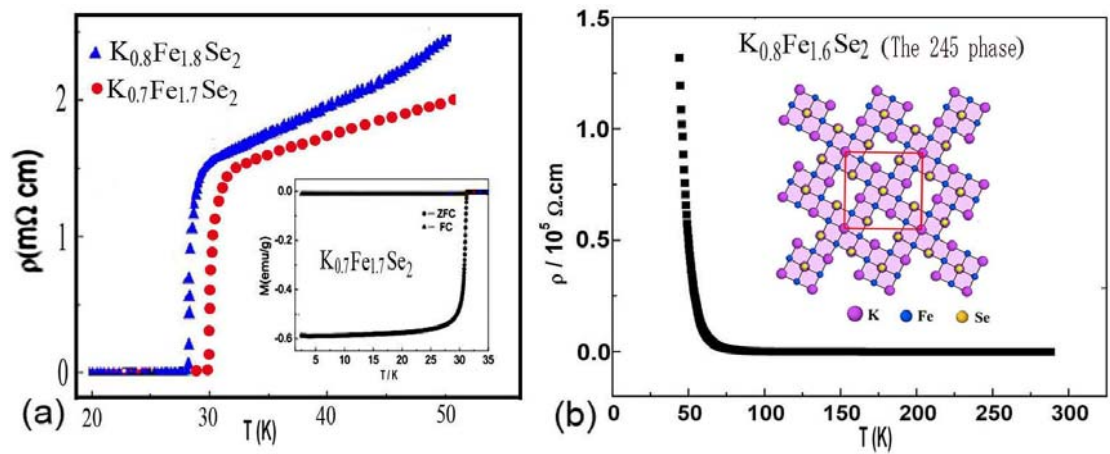

Fig.1

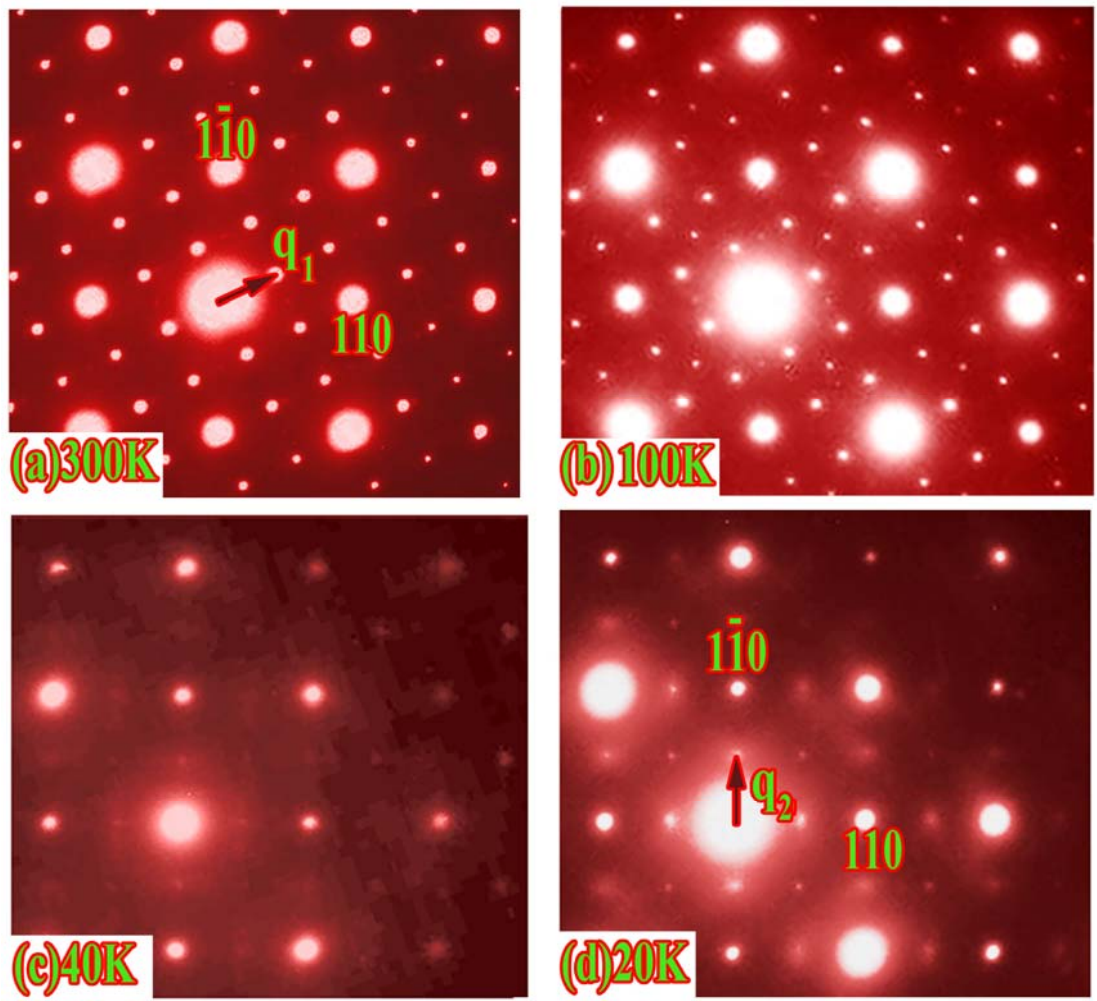

Fig.2

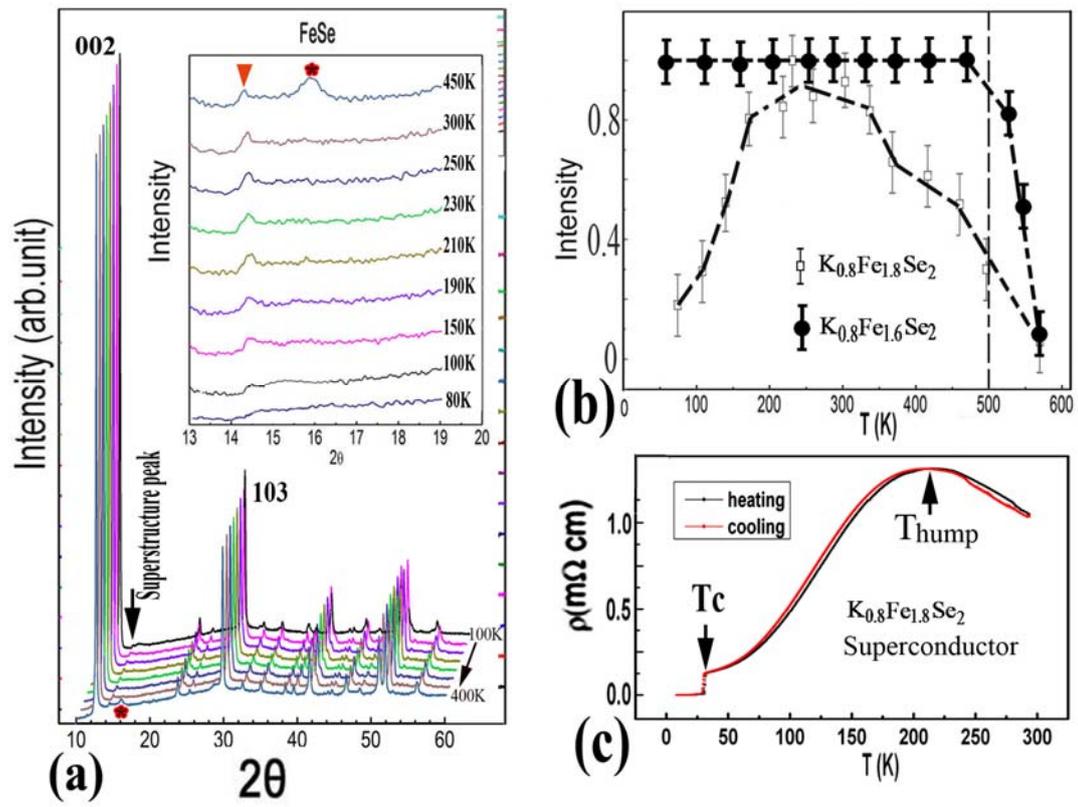

Fig.3

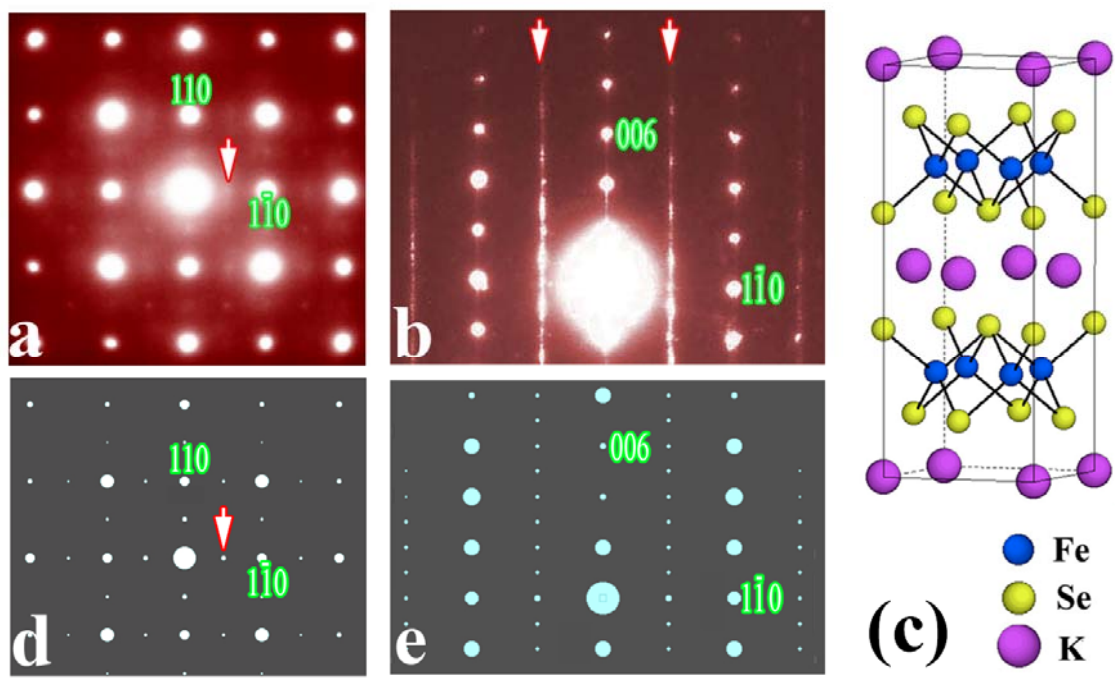

Fig.4